\documentclass{iau} 
\usepackage{graphicx}

\title[Star Formation Histories of the LG Dwarf Galaxies] 
{The Local Group Dwarf Galaxies. The Star Formation Histories derived using the Long Period Variable Stars}

\author[Elham Saremi et al.]   
{Elham Saremi$^1$, 
Atefeh Javadi$^1$, 
Jacco Th. van Loon$^2$,
Habib Khosroshahi$^1$, 
Sara Rezaeikh$^3$, 
Roya Hamedani Golshan$^4$, 
Seyed Azim Hashemi$^{5,1}$}

\affiliation{$^1$School of Astronomy, Institute for Research in Fundamental Sciences (IPM), Tehran, 19395-5531, Iran
\\ email: {\tt saremi@ipm.ir}    
\\[\affilskip]$^2$Lennard-Jones Laboratories, Keele University, ST5 5BG, UK
\\[\affilskip]$^3$Max-Planck-Institut f\"{u}r Astronomie, K\"{o}nigstuhl 17, D-69117 Heidelberg, Germany
\\[\affilskip]$^4$Physikalisches Institut der Universit\"{a}t zu K\"{o}ln, Z\"{u}lpicher Str. 77, D-50937 K\"{o}ln, Germany
\\[\affilskip]$^5$Physics Department, Sharif University of Technology, Tehran 1458889694, Iran}

\pubyear{2018}
\volume{344}  
\jname{	Dwarf Galaxies: From the Deep Universe to the Present}
\editors{A.C. Editor, B.D. Editor \& C.E. Editor, eds.}
\begin{document}

\maketitle

\begin{abstract}
Dwarf galaxies in the Local Group (LG) represent a distinct as well as diverse family of tracers of the earliest phases of galaxy assembly
and the processing resulting from galactic harassment. Their stellar populations can be resolved and used as probes of the evolution of
their host galaxy. In this regard, we present the first reconstruction of the star formation history (SFH) of them using the most evolved 
AGB stars that are long period variable (LPV). LPV stars trace stellar populations as young as $\sim$ 30 Myr to as old as the oldest globular
clusters. For the nearby, relatively massive and interacting gas-rich dwarf galaxies, the Magellanic Clouds, we found that the bulk
of the stars formed $\sim$ 10 Gyr ago for the LMC, while the strongest episode of star formation in the SMC occurred a few Gyr
later. A peak in star formation around 0.7 Gyr ago in both Clouds is likely linked to their recent interaction. The Andromeda satellite
pair NGC147/185 show different histories; the main epoch of star formation for NGC\,185 occurred 8.3 Gyr ago, followed by a much lower,
but relatively constant star formation rate (SFR). In the case of NGC\,147, the SFR peaked only 6.9 Gyr ago, staying intense until
$\sim$ 3 Gyr ago. Star formation in the isolated gas-rich dwarf galaxy IC\,1613 has proceeded at a steady rate over the past 5 Gyr,
without any particular dominant epoch. Due to lack of sufficient data, we have conducted an optical monitoring survey at the Isaac
Newton Telescope (INT) of 55 dwarf galaxies in the LG to reconstruct the SFH of them uniformly. The observations are made over ten
epochs, spaced approximately three months apart, as the luminosity of LPV stars varies on timescales of months to years. The system
of galactic satellites of the large Andromeda spiral galaxy (M31) forms one of the key targets of our monitoring survey. We present
the first results in the And I dwarf galaxy, where we discovered 116 LPVs among over 10,000 stars.
\keywords{stars: formation, stars: AGB, stars: variables: LPV, Galaxy: stellar content, (galaxies:) Local Group, galaxies: dwarf.}
\end{abstract}

\firstsection 
\section{Introduction}

One of the best ways for finding out more about the structure and evolution of galaxies is 
to study the star formation history (SFH) of the LG dwarf galaxies, due to their proximity, 
high number and variety. 
To investigate the SFH, we have developed a method based on employing LPV stars 
(\cite[Javadi et al. 2011b]{Javadi_etal11b};
\cite[Javadi et al. 2011c]{Javadi_etal11c};
\cite[Rezaeikh et al. 2014]{Rezaeikh_etal14}; 
\cite[Hamedani Golshan et al. 2017]{Golshan_etal17};
\cite[Javadi et al. 2017]{Javadi_etal17}),
which are mostly AGB stars at their very late stage of evolution (\cite[Javadi et al. 2011a]{Javadi_etal11a}; \cite[Javadi et al. 2015]{Javadi_etal15}).

\section{Methodology}

To describe the SFH, we identify the LPV stars in a galaxy and use their brightness
distribution function, $f(mag)$, to construct the birth mass function and hence 
the Star Formation Rate (SFR), $\xi$, as a function of look-back time (``age''), $t$: 
\begin{equation}
\xi(t) = \frac{f(mag(M(t)))}{\Delta(M(t))f_{\rm IMF}(M(t))},
\end{equation}
where $\Delta$ is the duration of the evolutionary phase during which LPV stars
display strong radial pulsation, and $f_{\rm IMF}$ is the Initial Mass
Function describing the relative contribution to star formation by stars of
different mass. Each of these functions depends on the stellar mass, $M$,
and the mass of a pulsating star at the end of its evolution is directly 
related to its age
(\cite[Javadi et al. 2011b]{Javadi_etal11b}).

\section{Data and Results}

\begin{figure}[b]
\begin{center}
 \includegraphics[width=2.7in]{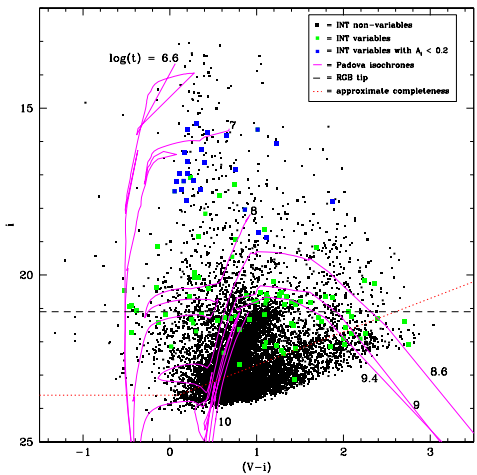} 
 \caption{The colour--magnitude diagram of $(V-i)$ for And\,I, with variable stars highlighted 
in green. The variable stars with $A_{\rm i}<0.2$ mag are highlighted in blue. 
Overplotted are isochrones from Marigo et al.\ (2008) with a distance modulus of 24.49 mag.}
\label{fig1}
\end{center}
\end{figure}

We are presenting the SFH of the dwarf galaxies in the LG uniformly 
based on identifying their LPV stars. In this regard, an optical long-term monitoring survey, 
with the Isaac Newton Telescope (INT), of the majority of dwarf galaxies in the LG are conducted
(\cite[Saremi et al. 2017a]{Saremi_etal17a}),
except those have sufficient data and/or are not visible in the Northern hemisphere survey
such as NGC\,147, NGC\,185, IC\,1613 and Magellanic Clouds (the LMC and SMC).

The first results of this survey in the And\,I dwarf galaxy show 116 LPVs among over 10,000 identified stars.
Fig.\,\ref{fig1} presents the colour--magnitude diagram of $(V-i)$ for And\,I, 
with LPV stars highlighted in green. The variable stars with $A_{\rm i}<0.2$ mag 
are highlighted in blue. Overplotted are isochrones from 
\cite[Marigo et al. (2008)]{Marigo08} 
with a distance modulus of 24.49 mag and Z = 0.00069 
(\cite[Kalirai et al. 2010]{Kalirai10}).
The 10 Gyr isochrone defines the location of tip of the RGB, that here is 21.1 mag for $i$ band.
The dotted line marks the 50\% completeness limit
(\cite[Saremi et al. 2017b]{Saremi_etal17b}).

\begin{figure}
\begin{center}
 \includegraphics[width=2.6in]{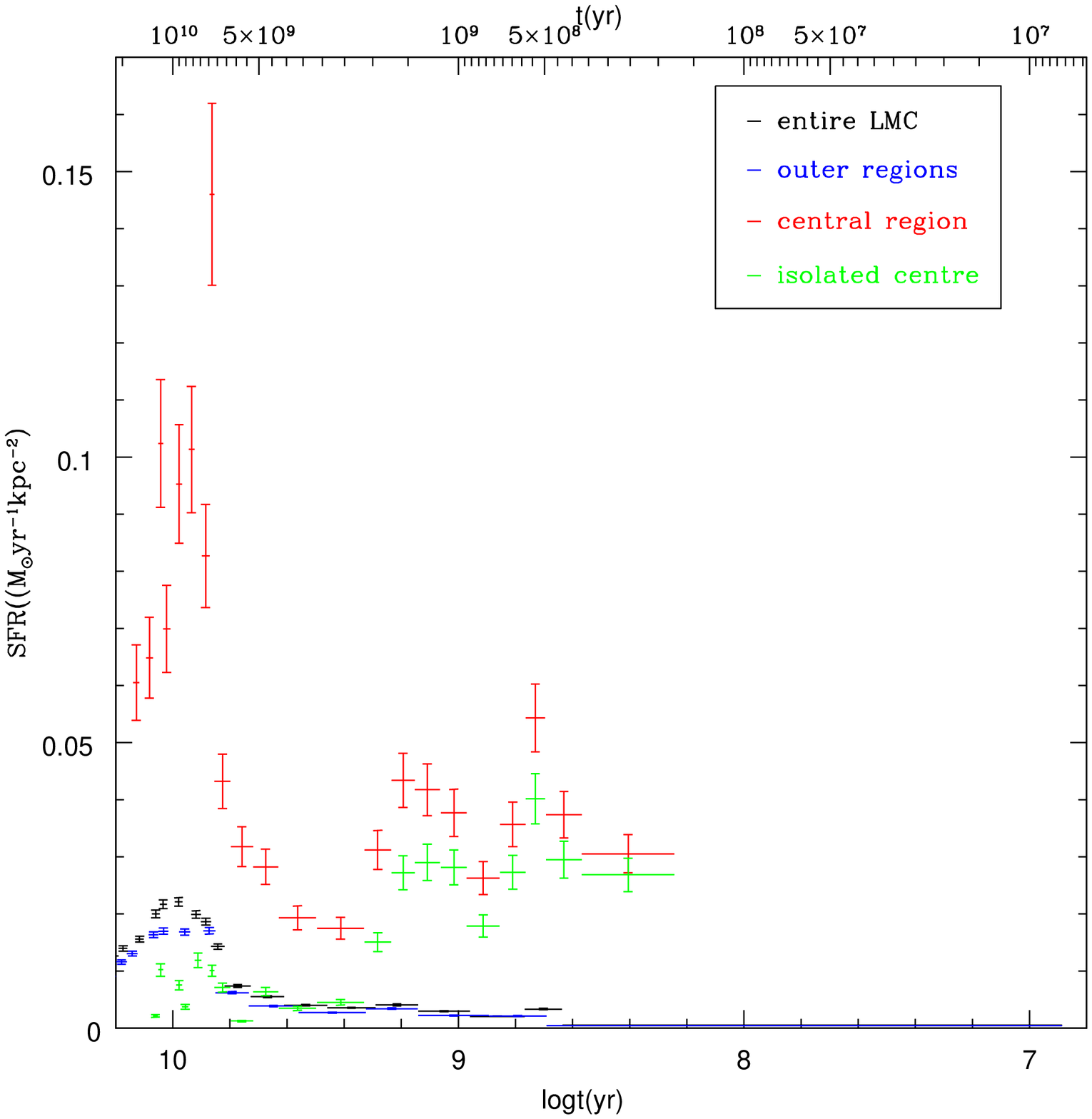}
 \includegraphics[width=2.6in]{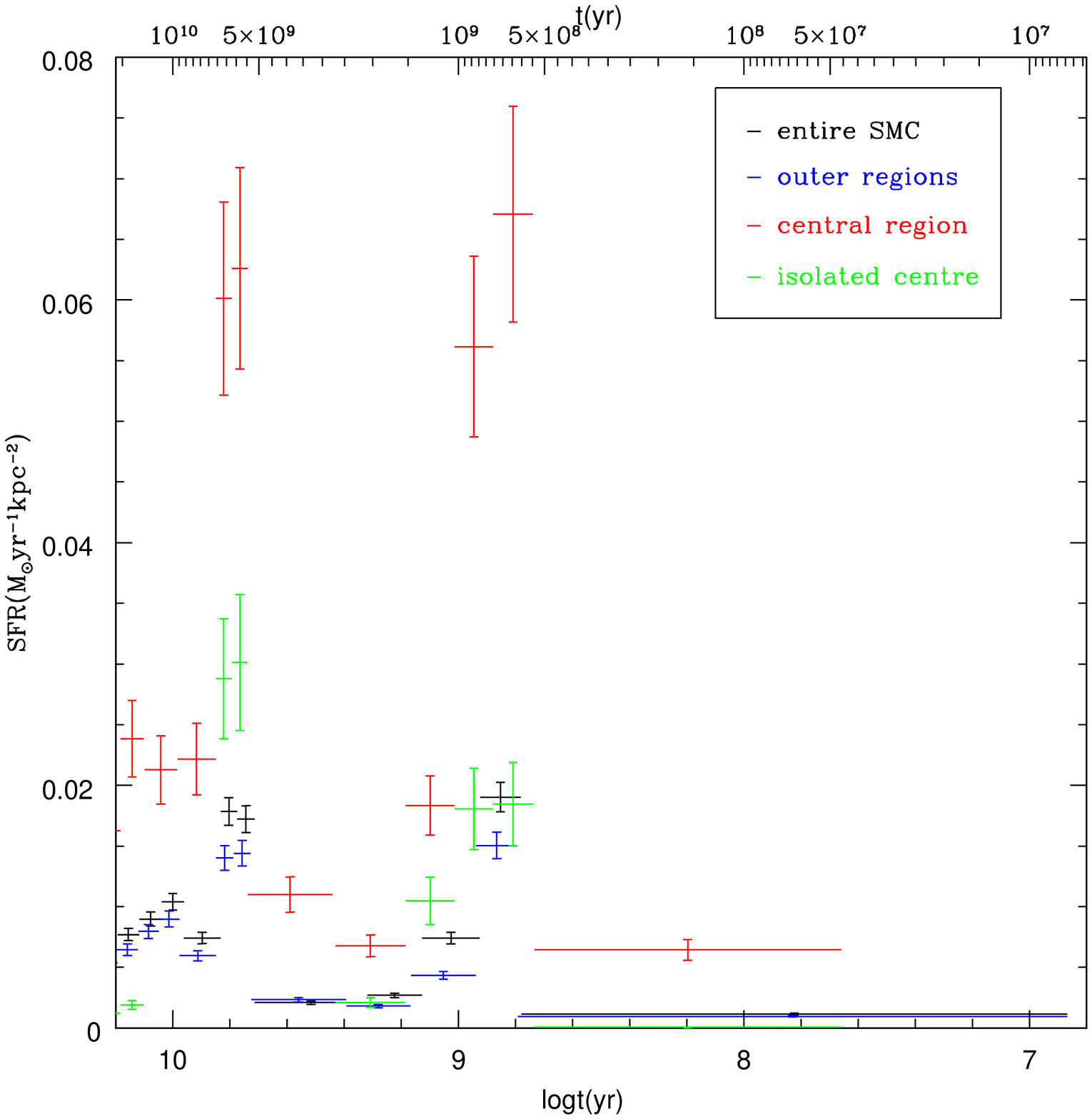} 
\caption[]{SFHs in the different regions of the LMC ({\it left}) and the SMC ({\it right}).
Black symbols: global star formation; blue symbols: star formation in the outskirt of the 
galaxies; red symbols: bar (for the LMC) and central (for the SMC) star formation; and 
green symbols: isolated star formation for the central regions derived by subtracting the 
SFH of the surrounding parts; The vertical lines show statistical error bars and the 
horizontal lines represent the age bins
(\cite[Rezaeikh et al. 2014]{Rezaeikh_etal14}).} 
  \label{fig2}
\end{center}
\end{figure}

Results for the Magellanic Clouds were obtained based on the variable star catalogues of 
\cite[Spano et al.\ (2011; LMC)]{Spano_etal11}
and
\cite[Soszy\'{n}ski et al.\ (2011; SMC)]{Soszy_etal11}
with 43,551 and 19,384 LPV stars, respectively. 
As is shown in Fig.\,\ref{fig2}, the bulk of the stars formed $\sim$ 10 Gyr ago for the LMC, 
at a rate of $1.598\pm0.054$ M$_\odot$ yr$^{-1}$, while for the SMC, two formation epochs at 
$\sim6$ Gyr, with SFR $0.282\pm0.017$ M$_\odot$ yr$^{-1}$ and $\sim0.7$ Gyr ago, with SFR 
$0.310\pm0.019$ M$_{\odot}$ yr$^{-1}$, are observed.
We derived the total stellar mass produced, $\sim2.2\times10^{10}$ M$_\odot$ for the LMC and 
$\sim4.0\times10^8$ M$_\odot$ for the SMC
(\cite[Rezaeikh et al. 2014]{Rezaeikh_etal14}).

\begin{figure}
\begin{center}
 \includegraphics[width=2.5in]{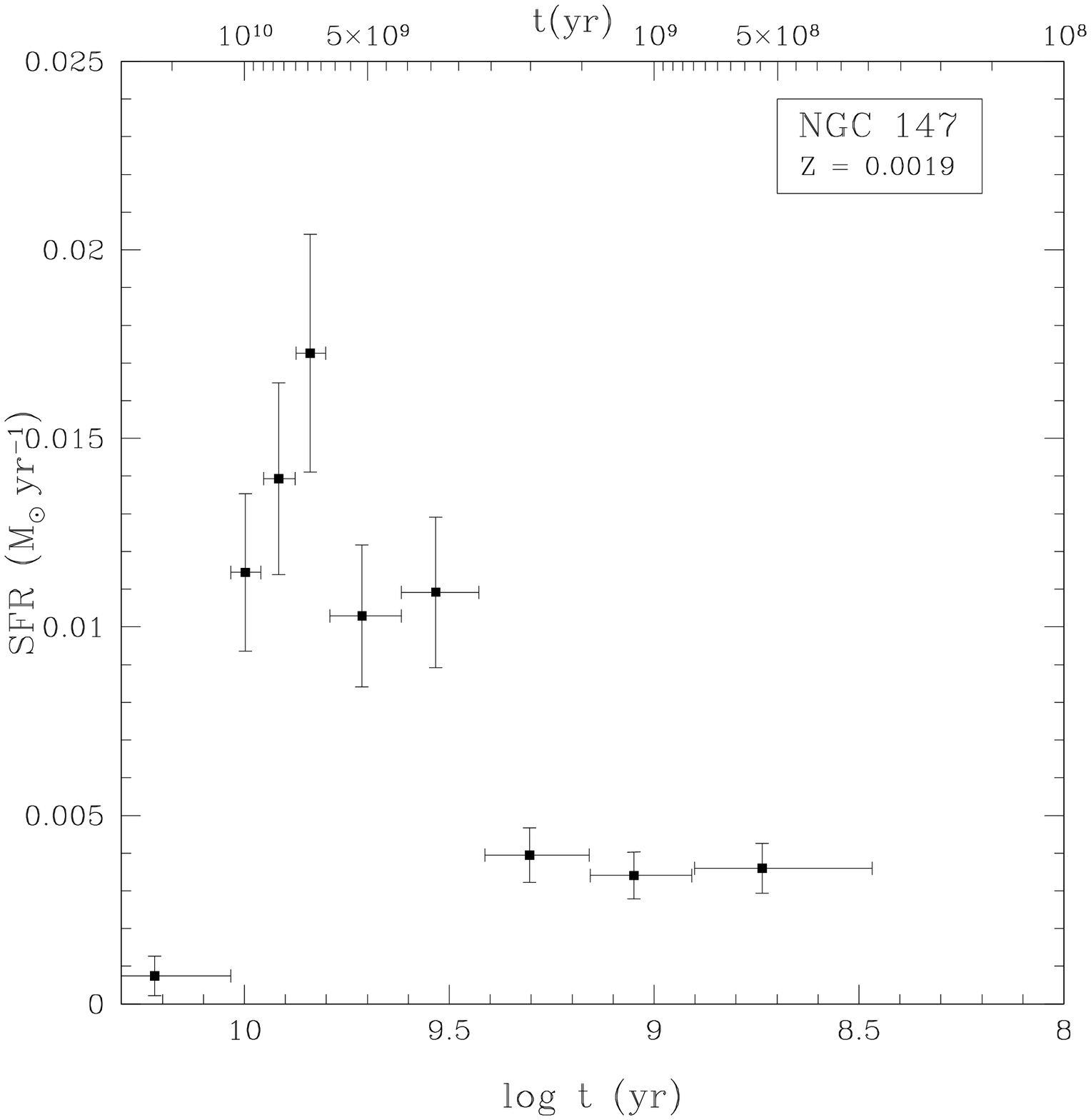}
 \includegraphics[width=2.5in]{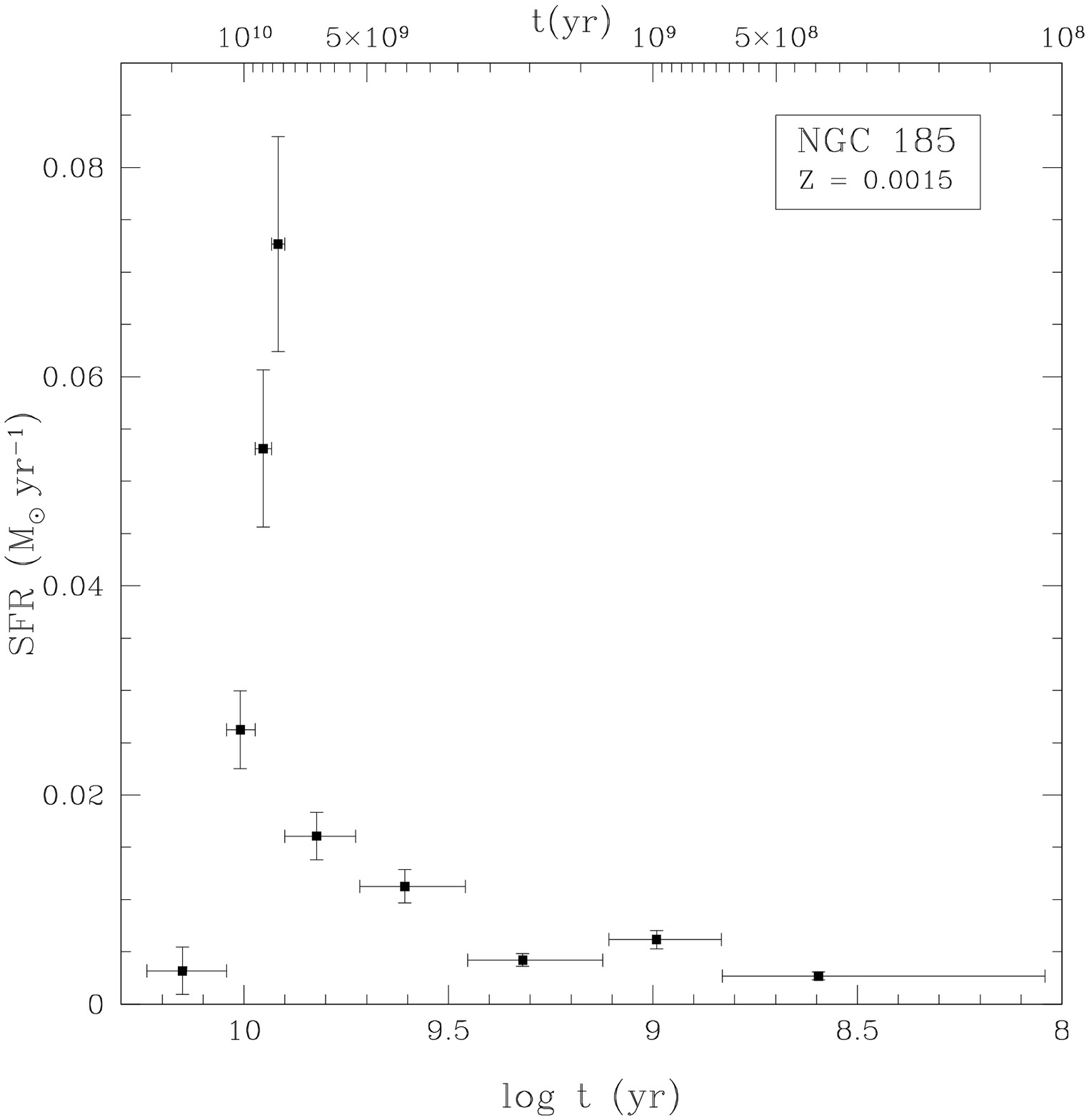}  
 \caption{SFHs in the central $6\rlap{.}^\prime5\times6\rlap{.}^\prime5$ regions
of NGC\,147 ({\it left}) and NGC\,185 ({\it right}), from LPV counts assuming
a constant metallicity 
(\cite[Hamedani Golshan et al. 2017]{Golshan_etal17}).}
   \label{fig3}
\end{center}
\end{figure}

NGC\,147 and NGC\,185 are two of the Andromeda satellites, which despite their similar mass 
and morphological type, show different SFHs. Using the catalogue of LPV stars from 
\cite[Lorenz et al.\ 2011]{Lorenz_etal11}
we estimated that the star formation started earlier in NGC\,185 (8.3 Gyr ago) than in NGC\,147 
(6.9 Gyr ago) and continued over the past 6 Gyr until 100 Myr ago, albeit at a much lower rate, 
while no star formation is seen for the past 300 Myr in NGC\,147 (Fig.\,\ref{fig3}).
We calculated the total stellar masses of $M\approx1.13\times10^8$ M$_\odot$ and 
$M\approx2.3\times10^8$ M$_\odot$ in NGC\,147 and NGC\,185, respectively
(\cite[Hamedani Golshan et al. 2017]{Golshan_etal17}).

\begin{figure}
\begin{center}
 \includegraphics[width=2.9in]{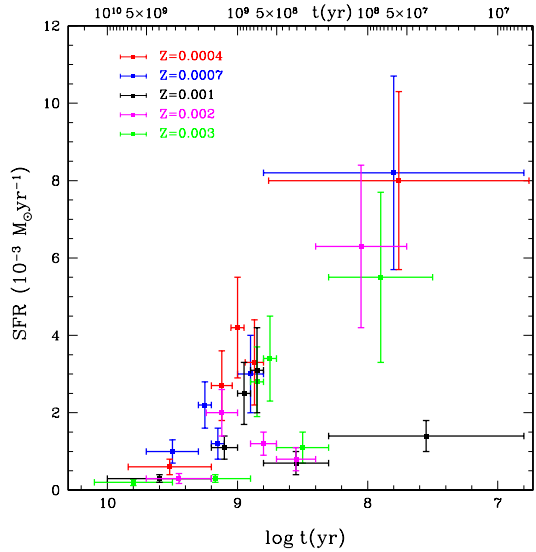} 
 \caption{IC\,1613 SFH for different metallicities
(\cite[Hashemi et al. 2019]{Hashemi_etal19}).}
   \label{fig4}
\end{center}
\end{figure}

Also, we studied the SFH of the isolated gas-rich dwarf galaxy IC\,1613, 
based on identified LPV stars by
\cite[Menzies et al. (2015)]{Menzies_etal15} 
and
\cite[Boyer et al. (2015)]{Boyer_etal15}.
Fig.\,\ref{fig4} shows the SFH of IC\,1613 for different metallicities. 
With adopting the amount Z = 0.003 for last Gyr ($log\,t(yr)<9$), 
Z = 0.002 for $1\, Gyr< t < 2\, Gyr$ ($9<log\,t(yr)<9.3$), and Z = 0.0007 
for $t>2\,Gyr$ ($log\,t(yr)>9.3$), it can be concluded that there is not 
any dominant episode of star formation in IC\,1613 over the past 5 Gyr
(\cite[Hashemi et al. 2019]{Hashemi_etal19}). 

\begin{thebibliography}{}

\bibitem[Boyer et al. (2015)]{Boyer_etal15}
{Boyer, M.L., McQuinn, K.B.W., Barmby, P., et al.} 2015,
\textit{ApJS}, 216, 10

\bibitem[Golshan et al. (2017)]{Golshan_etal17}
{Hamedani Golshan, R., Javadi, A., van Loon, J.Th., Khosroshahi, H., \& Saremi, E.} 2017, 
\textit{MNRAS}, 466, 1764

\bibitem[Hashemi et al. (2018)]{Hashemi_etal19}
{Hashemi, S.A., Javadi, A., van Loon, J.Th.} 2019, 
\textit{MNRAS}, 483, 4751

\bibitem[Javadi et al. (2011a)]{Javadi_etal11a}
{Javadi, A., van Loon, J.Th., \& Mirtorabi, M.T.} 2011a, 
\textit{MNRAS}, 411, 263

\bibitem[Javadi et al. (2011a)]{Javadi_etal11b}
{Javadi, A., van Loon, J.Th., \& Mirtorabi, M.T.} 2011b, 
\textit{MNRAS}, 414, 3394

\bibitem[Javadi et al. (2011b)]{Javadi_etal11c}
{Javadi, A., van Loon, J.Th., \& Mirtorabi, M.T.} 2011c, 
\textit{ASPC}, 445, 497

\bibitem[Javadi et al. (2015)]{Javadi_etal15}
{Javadi, A., Saberi, M., van Loon, J.Th., Khosroshahi, H., Golabatooni N., \& Mirtorabi, M.T.} 2015, 
\textit{MNRAS}, 447, 3973

\bibitem[Javadi et al. (2017)]{Javadi_etal17}
{Javadi, A., van Loon, J.Th., Khosroshahi, H., Tabatabaei, F., Hamedani Golshan, R., \& Rashidi, M.} 2017, 
\textit{MNRAS}, 464, 2103

\bibitem[Kalirai et al. (2010)]{Kalirai_etal10}
{Kalirai, J.S., Beaton, R.L., Geha, M.C., et al.} 2010, 
\textit{ApJ}, 711, 671

\bibitem[Lorenz et al. (2011)]{Lorenz_etal11}
{Lorenz, D., Lebzelter, T., Nowotny, W., et al.} 2011, 
\textit{$A\&A$}, 532, 78

\bibitem[Marigo et al. (2008)]{Marigo_etal08}
{Marigo, P., Girardi, L., Bressan, A., Groenewegen, M.A.T., Silva, L., \& Granato, G.L.} 2014, 
\textit{$A\&A$}, 482, 883 

\bibitem[Menzies et al. (2015)]{Menzies_etal15}
{Menzies, J.W., Whitelock, P.A., \& Feast, M.W.} 2015, 
\textit{MNRAS}, 452, 910

\bibitem[Rezaeikh et al. (2014)]{Rezaeikh_etal14}
{Rezaeikh, S., Javadi, A., Khosroshahi, H., \& van Loon, J.Th.} 2014, 
\textit{MNRAS}, 445, 2214

\bibitem[Saremi et al. (2017a)]{Saremi_etal17a}
{Saremi, E., Javadi, A., van Loon, J.Th., et al.} 2017a, 
\textit{J. Phys. Conf. Ser}, 869, 012068

\bibitem[Saremi et al. (2017b)]{Saremi_etal17b}
{Saremi, E., Abedi, A., Javadi, A., van Loon, J.Th., \& Khosroshahi, H.} 2017b, 
\textit{IJAA}, 4, 19

\bibitem[Soszy\'{n}ski et al. (2011)]{Soszy_etal11}
{Soszy\'{n}ski, I., Udalski, A., Szyma\'{n}ski, M.K., et al.} 2011,
\textit{AcA}, 61, 217S

\bibitem[Spano et al. (2011)]{Spano_etal11}
{Spano, M., Mowlavi, N., Eyer, L., et al.} 2011,
\textit{$A\&A$}, 536, 17

\end{thebibliography}
\end{document}